# INELASTIC SPIN-WAVE SCATTERING BY BLOCH DOMAIN WALL FLEXURE OSCILLATIONS

*Nataliya N. Dadoenkova, Yuliya S. Dadoenkova, Igor L. Lyubchanskii, Maciej Krawczyk, and Konstantin Y. Guslienko\**

**Prof. N. N. Dadoenkova, Dr. Y. S. Dadoenkova**
Ulyanovsk State University,
Ulyanovsk, 432000, Russia
Donetsk Institute for Physics and Engineering of the National Academy of Sciences of Ukraine, Ukraine
**Prof. I. L. Lyubchanskii**
Donetsk Institute for Physics and Engineering of the National Academy of Sciences of Ukraine, Ukraine
Faculty of Physics, V. N. Karazin Kharkiv National University,
Kharkiv, 61000, Ukraine
**Prof. M. Krawczyk**
Faculty of Physics, Adam Mickiewicz University,
Poznan, 61614, Poland
**Prof. K. Y. Guslienko**
Depto. Física de Materiales, Universidad del País Vasco, UPV/EHU,
Donostia, 20018, Spain
IKERBASQUE, The Basque Foundation for Science,
Bilbao, 48013, Spain
*E-mail: kostyantyn.gusliyenko@ehu.eus*

The calculations of the inelastic spin wave scattering by flexure vibrations of the Bloch domain wall (Winter´s magnons) in thin magnetic films are presented. The approach is based on the interaction of the propagating spin waves with the dynamical emergent electromagnetic field generated by the moving inhomogeneous magnetization texture (domain wall). The probability of the spin wave scattering for the Winter´s magnon emission and absorption processes essentially rises with the spin wave scattering angle increase up to $90^0$. The angular dependence of the scattering probability is essentially stronger for the magnon absorption processes that allow distinguishing these elementary emission/absorption processes experimentally.

Spin wave (SW)[1] propagation in inhomogeneously magnetized magnetic films, magnetic stripes and artificial periodic nanostructures (magnonic crystals)[2] is the subject of intensive theoretical and experimental investigations because of promising applications of such magnetic structures in spintronic and magnonic devices.[3, 4] Usually, the propagation of SW through static and/or pinned domains [5-9] and SW emission by oscillating domain wall (DW)[10, 11] is studied. However, it is well known that some flexure oscillations of the DW shape exist in DW (so-called Winter´s magnons), and they can be represented as specific SW localized at the DW plane.[12-14] These magnetization oscillations propagate along the DW plane and can modulate incident waves of different nature (electromagnetic waves[15, 16], microwave phonons[17], neutrons[18]) in a magnetic medium with DWs. The localized Winter´s SWs were observed in the double-vortex state magnetic dots as oscillations of the Néel DW connecting the vortex cores[19]. They also were simulated[20] and observed[21] in magnetic stripes, and named as SW "nano-channeling".

Several research groups considered recently the influence of the SW propagating in nanostripes on DW displacement as a whole. It was demonstrated that the DW motion is strongly dependent on whether the SWs pass through the DW or they are reflected by the wall. In the first case, DW propagates in the opposite direction to that of the incident SW[22]. When SW is partially reflected by the DW, the momentum transfer between the SW and DW results in a torque, which drives the DW to move in the same direction as the incident SW[23]. These effects are valid within the linear SW theory and neglecting internal degrees of freedom of DW (assuming a rigid DW profile). The Landau-Lifshitz equation of the magnetization motion within the linear theory can be reduced to the Schrödinger equation with a static potential determined by the DW profile. For some specific kind of the potential (Pösch-Teller potential[24]) created by 1D static DW, this equation has solutions corresponding to the SW propagation trough the DW without reflection.

However, there are some internal DW oscillation modes which can interact with SWs resulting in an inelastic scattering of SWs by the DW modes or excitations of the DW modes by SWs. This effect is analogous to the inelastic, or Brillouin light scattering (BLS) by SWs with replacing the light by SWs and SWs by DW excitations.[25] These inelastic scattering processes are characterized in the main approximation by three magnetic excitation modes interaction (it is cubic in the mode amplitudes) and were described, for example, in the book[26] and papers.[27, 28] In several papers [5-7] it is stated that the SW travelling in a magnetic stripe along its length excites DW translation motion (DW velocity peaks at the quantized SW frequencies) due to direct processes, which are described by the interaction term which is bi-linear in the SW and DW amplitudes. For the small stripe width $L_y$ =50 nm [5] or 150 nm [6] it is necessary to account the quantization of the transverse wave vector $\kappa_y^n \approx \pi n / L_y$ ($n$ =1, 3,…). In this case the DW frequencies $\Omega_n$ can be large, 18-27 GHz ($n$ =3, 5) [5] or 8-11 GHz (n=3) [6], comparable with the delocalized SW frequencies and direct SW – Winter´s magnon scattering is allowed [5-7]. Below we consider the case of a thin infinite film, where the lateral quantization of the DW wave vector is irrelevant and the DW excitation frequencies are essentially smaller than typical SW frequencies (10-20 GHz), and, therefore, the direct SW-DW scattering processes are prohibited due to the energy conservation.

In this article, we consider the scattering of SWs by DW oscillations in a magnetic film with a single Bloch DW. We investigate an angular dependence of the scattered SW modulated by the DW oscillations and present analytical calculations of the expected effects.

Let us consider a thin film with two magnetic domains with opposite directions of magnetization separated by a Bloch DW as depicted in **Figure 1**. There are two kinds of spin excitations on the DW background. SWs (localized far from the DW plane) and flexure oscillations of the DW shape (Winter´s magnons). We denote the amplitudes of the DW excitations as $a_n$, and SW excitations as $b_k$, where the indices $n$ and $k$ are numbering the modes. We describe the unit magnetization vector **m**($\Theta$, $\Phi$) = **M**/$M_s$ by the spherical angles $\Theta$ and $\Phi$. The SWs are defined as small perturbations on the DW background (dynamical, in the general case). We consider a Bloch DW with its plane $yOz$ and the $x$-axis perpendicular to the wall plane. The components of the SW magnetization **m**$_s$ are the simplest in a moving coordinate frame $x'y'z'$, $\mathbf{m}'_s = (\vartheta, \sin\Theta_\upsilon \psi, 0)$, where the axis $Oz'$ is directed along the local instant direction of the DW magnetization defined by the angles $(\Theta_\upsilon, \Phi_\upsilon)$, and the SW angles $\vartheta, \psi$ are small deviations from the local DW angles $\Theta = \Theta_\upsilon + \vartheta$, $\Phi = \Phi_\upsilon + \psi$. From the other side, there are small excitations of the static DW background described as $\Theta_\upsilon = \Theta_0 + \eta$, $\Phi_\upsilon = \Phi_0 + \zeta$. The inelastic scattering of SWs by the DW oscillations is determined by an interaction Hamiltonian $H_{int}$, which is deduced below. It contains bilinear (*ab*), cubic (*aab*), (*abb*), quartic (*aabb*), and other high order interaction terms.

It was shown[30] that, in general, SW on the background of a magnetic soliton can be described by the Lagrangian where the derivatives $\partial_\mu$ are substituted by the covariant derivatives $D_\mu = (\partial_\mu - \hat{A}_\mu)$, $x_\mu$ = *ct, x, y, z*, with *c* being the speed of light in vacuum, and $\partial_\mu = \partial/\partial x_\mu$. Here $\hat{A}_\mu = \partial_\mu R \cdot R^{-1}$ is pure gauge potential related to the transformation to the local coordinate frame connected to the soliton magnetization **m**′ = *R***m**, *R* is the rotation matrix. The 3x3 matrix $\hat{A}_\mu$ allows to define a dual vector $\mathbf{A}_\mu$ by the relation $\hat{A}_\mu \mathbf{m} = \mathbf{m} \times \mathbf{A}_\mu$. The vector potential of the emergent electromagnetic field $A_\mu^e = (\mathbf{A}_\mu \cdot \mathbf{m}')$ is responsible for an interaction of the slowly moving inhomogeneous magnetization background (DW) and SWs.[30] The bilinear (*ab*) and cubic (*aab*) SW-DW interaction terms come from kinetic part of the Lagrangian density $L_{kin} = (M_s/\gamma)(1-\cos\Theta)\dot{\varsigma}$ [30]

$$L_{int} = \frac{M_s}{\gamma}\sin\Theta_0 \left(\dot{\varsigma} \quad \dot{} \quad \dot{} \quad \right) \tag{1}$$

where $M_s$ is the material saturation magnetization, $\gamma$ is the gyromagnetic ratio, and overdot marks a derivative with respect to time.

However, the first (*ab*)-terms assume some direct processes involving creation/annihilation of SW and annihilation/creation of a Winter's magnon. Such processes are prohibited because of the DW oscillation frequencies usually are much smaller than the SW frequencies. The second term in Equation (1) describing (*aab*)-processes is small because the function determining the static wall profile $\cos\Theta_0(x) = \tanh(x/\Delta)$ ($\Delta$ is the DW thickness) is small near the DW plane *x* = 0, where the DW oscillations $\eta, \varsigma$ are localized. The SW-DW interaction terms coming from $L_{kin}$ were analyzed in details by Le Maho et al.[31] Other SW-DW interaction terms appear due to the exchange energy in the Lagrangian $L = L_{kin} - w$, where the energy density *w* includes contributions from the exchange interaction, magnetic anisotropy, dipolar interaction, *etc*. The SW exchange energy density is

$$w_{ex}(\mathbf{m}_s) = A(D_a\mathbf{m}_s)^2 = A(\partial_a\mathbf{m}_s + A_a^e\mathbf{m}\times\mathbf{m}_s)^2 = A\left[(\nabla + i\mathbf{A}^e)\Psi\right]^2, \tag{2}$$

where *A* is the exchange stiffness, $\mathbf{A}^e = (1-\cos\Theta_\upsilon)\nabla\Phi_\upsilon$ is the spatial part of the Abelian gauge potential $A_\mu^e = (1-\cos\Theta_\upsilon)\partial_\mu\Phi_\upsilon$, *a* = *x,y,z*, and the vector $\mathbf{m}_s$, complex variable $\Psi = \vartheta + i\sin\Theta_\upsilon\psi$ describe the SW magnetization.

The other contributions to the magnetic energy density (magnetic anisotropy, magnetostatic energy, etc.) do not contain spatial derivatives of the magnetization and, therefore, do not contribute to the DW-SW interaction. It is convenient to rewrite the SW-DW interaction part $w_{int}$ in Equation (2) as

$$w_{int} = A\left[-\mathbf{j}\cdot\mathbf{A}^e + (\mathbf{A}^e)^2|\Psi|^2\right], \tag{3}$$

where the spin current **j** is determined as $\mathbf{j} = i\left(\Psi^*\nabla\Psi - \Psi\nabla\Psi^*\right)$.

In our case of the incident (i) and scattered (s) SWs, the magnon's wave function $\Psi$ can be presented as a linear combination $\Psi = \Psi_i + \Psi_s$. In the expression for the spin current we take into account only the cross-terms which correspond to the SW scattering process $\delta\mathbf{j}$, namely, $\delta\mathbf{j} = i\left(\Psi_i^*\nabla\Psi_s + \Psi_s^*\nabla\Psi_i - \Psi_i\nabla\Psi_s^* - \Psi_s\nabla\Psi_i^*\right)$.

The emergent field vector potential $\mathbf{A}^e$ created by the moving DW can be separated in two parts, static and dynamic ones, $\mathbf{A}^e = \mathbf{A}_0^e + \mathbf{a}^e$, $\mathbf{A}_0^e = (1 - \cos\Theta_0)\nabla\Phi_0$ (equal to zero for the partial case of the Bloch DW having the static angles $\Phi_0 = \pm\pi/2$), and $\mathbf{a}^e = (1 - \cos\Theta_0)\nabla\zeta$. Therefore, the cubic (*abb*) SW-DW interaction terms can be written in the form

$$w_{\text{int}} = -A(1 - \cos\Theta_0)\mathbf{j}\cdot\nabla\zeta. \tag{4}$$

The incident (i) and scattered (s) SWs can be decomposed in series of the eigenmodes,

$$\Psi_{(\alpha)}(\mathbf{r},t) = \sum_{\mathbf{k}} b_{\mathbf{k}}^{(\alpha)} m_{\mathbf{k}}(\mathbf{r})\exp(i\omega_\alpha t), \tag{5}$$

where $\alpha = (i, s)$, $b_{\mathbf{k}}^{(\alpha)}$ is the SW eigenmode amplitude, $m_{\mathbf{k}}(\mathbf{r})$ is the SW eigenmode profile, and $\omega_\alpha$ is the eigenmode frequency.

The spin current density for the case of a single plane wave with the wave vector **k**, $m_{\mathbf{k}}(\mathbf{r}) = \exp(-i\mathbf{k}\mathbf{r})$, is $\mathbf{j} = 2|b_{\mathbf{k}}|^2\mathbf{k}$. Therefore, the first term in Equation (4) differs from zero only if one of the i-, s-SWs has a non-zero wave vector component parallel to the DW plane. Let us denote the wave vectors and frequencies of the incident and scattered SW as $(\mathbf{k}, \omega_{\mathbf{k}})$ and $(\mathbf{K}, \omega_{\mathbf{K}})$, respectively. Substituting Equation (5) into the equation for the spin current $\delta\mathbf{j}$ we obtain

$$\delta\mathbf{j} = -2\,\mathrm{Im}\sum_{\mathbf{K},\mathbf{k}}\left[b_{\mathbf{K}}^* b_{\mathbf{k}}\left(m_{\mathbf{K}}^*\nabla m_{\mathbf{k}} - m_{\mathbf{k}}\nabla m_{\mathbf{K}}^*\right)e^{i(\omega_{\mathbf{k}} - \omega_{\mathbf{K}})t}\right]. \tag{6}$$

It is convenient to consider that the SW wave vector components are discrete (quasi-continuous). The total number of magnons in the system is $V_0^{-1}\int d^3\mathbf{r}\,\Psi^*\Psi = \sum_{\mathbf{k}} b_{\mathbf{k}}^* b_{\mathbf{k}}$ ($V_0 = a_0^3$ is the volume per unit atomic spin, $a_0$ is the lattice period, and $\delta_{\mathbf{kk'}}$ is the Kronecker symbol). This leads to the natural normalization of the magnon eigenmodes $m_{\mathbf{k}}(\mathbf{r})$ presented in Equation (5), $V_0^{-1}\int d^3\mathbf{r}\,m_{\mathbf{k}}^* m_{\mathbf{k'}} = \delta_{\mathbf{kk'}}$. In the case of the uniaxial magnetic anisotropy and 1D Bloch or Neel DW (the walls are described by $\Theta_0(0) = \pi/2$, the DW plane is *yOz*) the normalized SW mode profile is $m_{\mathbf{k}}(\mathbf{r}) = \left[\tanh(x/\Delta) + ik_x\Delta\right]\exp\left[-i(\mathbf{k}\cdot\mathbf{r})\right]/\left(\sqrt{N(1 + k_x^2\Delta^2)}\right)$,[12] where $N = V/V_0$ is the number of the spins in the sample. Calculating the gradient terms in Equation (6) we get the expression

$$m_{\mathbf{K}}^{*}\nabla m_{\mathbf{k}}-m_{\mathbf{k}}\nabla m_{\mathbf{K}}^{*}=\frac{\left[-i(\mathbf{k}+\mathbf{K})(\cos\Theta_{0}-iK_{x}\Delta)(\cos\Theta_{0}+ik_{x}\Delta)+i(K_{x}+k_{x})\Delta\hat{\mathbf{x}}\sin\Theta_{0}\Theta_{0}'\right]}{N\sqrt{1+K_{x}^{2}\Delta^{2}}\sqrt{1+k_{x}^{2}\Delta^{2}}}e^{i\mathbf{kr}}.$$

(6′)

The dynamic DW magnetization can be written in the complex form $\chi=\eta+i\sin\Theta_{0}\zeta$. The decomposition of $\chi$ via the DW normal modes yields

$$\chi(\mathbf{r},t)=\sum_{\kappa}a_{\kappa}\mu_{\kappa}(\mathbf{r})\exp\left[i(\Omega_{\kappa}t)\right], \tag{7}$$

where $\boldsymbol{\kappa}$ is the wave vector in the DW-plane $yOz$, $\Omega_{\kappa}$ is the DW oscillation frequency, and $\mu_{\kappa}(\mathbf{r})=(N_{w})^{-1/2}\operatorname{sech}(x/\Delta)\exp\left[-i(\boldsymbol{\kappa}\cdot\mathbf{r})\right]$ is the normalized DW eigenmode profile, where $N_{w}=V_{DW}/V_{0}$ is the number of spins in the domain wall, $V_{DW}=2\Delta S$ is the effective DW volume, and $S$ is the DW surface square. The expression for the DW dynamical phase $\zeta$ is determined as $\zeta=(\sin\Theta_{0})^{-1}\operatorname{Im}(\chi)$, $\sin\Theta_{0}(x)=\operatorname{sech}(x/\Delta)$, and therefore,

$$\nabla\zeta=-\frac{1}{2N_{w}^{1/2}}\sum_{\kappa}\left[a_{\kappa}e^{i\Omega_{\kappa}t}-a_{-\kappa}^{*}e^{-i\Omega_{\kappa}t}\right]\boldsymbol{\kappa}e^{i\mathbf{kr}}. \tag{8}$$

We assume that the magnetic element (film) is thin and there is no dependence of the dynamical magnetization on the thickness coordinate $y$. Therefore, the wave vectors $\mathbf{k}$, $\mathbf{K}$, and $\boldsymbol{\kappa}$ have no $y$-component. Substituting Equations (6) and (8) to Equation (4) and integrating the interaction energy density (4) over the sample volume, we get the interaction Hamiltonian $H_{\text{int}}$ represented in terms of the DW and SW oscillation amplitudes in the following form

$$H_{\text{int}}=\sum_{\mathbf{k},\mathbf{K}}B(\mathbf{K},\mathbf{k})b_{\mathbf{K}}^{*}b_{\mathbf{k}}\left(a_{\kappa}e^{i\Omega_{\kappa}t}-a_{-\kappa}^{+}e^{-i\Omega_{\kappa}t}\right)e^{i(\omega_{\mathbf{k}}-\omega_{\mathbf{K}})t}, \tag{9}$$

where the scattering amplitude accounting the Bloch DW profile is

$$B(\mathbf{K},\mathbf{k})=A\frac{\Delta V_{0}}{L_{x}\sqrt{N_{w}}}\frac{\left(K_{y}^{2}-k_{y}^{2}\right)}{\sqrt{\left(1+K_{x}^{2}\Delta^{2}\right)\left(1+k_{x}^{2}\Delta^{2}\right)}}F(K_{x},k_{x}), \tag{10}$$

where $L_x$ is the magnetic element length in the $x$-direction, and the function $F$ is determined by the expression

$$F(K,k)=\frac{L_{x}}{\Delta}(1+Kk\Delta^{2})\delta_{q0}-\left(1-\frac{i}{2}\xi\right)\frac{\pi\xi}{\sinh\left(\frac{\pi}{2}\xi\right)}+\left[\xi+i(1+Kk\Delta^{2})\right]I(\xi), \tag{11}$$

with $\xi=q\Delta$, $q=k_{x}-K_{x}$, and the function $I(\xi)=\pi/\sinh(\pi\xi/2)-(2/\xi)\cos(L_{x}\xi/2\Delta)$. We can substitute the complex SW amplitudes $b_{\mathbf{k}}^{*},b_{\mathbf{k}}$ in the Hamiltonian (9) to the Bose creation/annihilation operators $b_{\mathbf{k}}^{+},b_{\mathbf{k}}$ and consider the elementary processes of the SW scattering described by the terms $B(\mathbf{K},\mathbf{k})b_{\mathbf{K}}^{+}b_{\mathbf{k}}a_{\kappa}$ and $B(\mathbf{K},\mathbf{k})b_{\mathbf{K}}^{+}b_{\mathbf{k}}a_{-\kappa}^{+}$ (see **Figure 2**). The probability of the SW scattering by DW oscillations per unit of time can be then presented within the first-order time-dependent perturbation theory[32] as

$$W(\mathbf{k} \to \mathbf{K}) = \frac{2\pi}{\hbar} |\langle \mathit{out} | \mathit{I}_{int} | \mathit{in} \rangle|^2 \delta(\omega_\mathbf{k} - \omega_\mathbf{K} \pm \Omega_\kappa). \tag{12}$$

There are in the initial state SW ($\mathbf{k}$, $\omega_\mathbf{k}$) and in the final state SW ($\mathbf{K}$, $\omega_\mathbf{K}$) and DW oscillation ($\kappa$, $\Omega_\kappa$) in the emission process (a) in Figure 2. In the absorption process shown in Figure 2 (b), the SW ($\mathbf{k}$, $\omega_\mathbf{k}$) oscillations and DW oscillation ($\kappa$, $\Omega_\kappa$) are in the initial state, while the SW ($\mathbf{K}$, $\omega_\mathbf{K}$) is in the final state. These processes are analogical to anti-Stokes and Stokes components in the Raman (or Brillouin) light scattering in condensed matter.[25, 33] There are a momentum conservation law $\mathbf{k} = \mathbf{K} \pm \kappa$ and energy conservation law $\omega_\mathbf{k} = \omega_\mathbf{K} \pm \Omega_\kappa$ for the emission (+) and absorption (−) processes, respectively.

We consider the SWs with wave lengths much larger than the DW width $\Delta \approx 10$ nm. Therefore, $k\Delta \ll 1$, $K\Delta \ll 1$, $\xi \ll 1$ and Equations (10) and (11) can be simplified. The function $F(K_x,k_x)$ in Equation (11) takes the form $F(K_x,k_x) = F(\xi) = -2\cos(\xi L_x / 2\Delta) + iI(\xi)$, and $|F(\xi)|^2 = 4\cos^2(\xi L_x / 2\Delta) + |I(\xi)|^2$. The function $I(\xi)$ reveals a sharp maximum $I(\xi_m) = 2L_x / \pi\Delta$ at $\xi_m \approx \pi\Delta / L_x$ due to the relation $L_x / 2\Delta \gg 1$ and goes to zero at $\xi \gg 1$. Therefore, we use $|F(\xi_m)|^2$ to calculate the scattering amplitude $|B(\mathbf{K},\mathbf{k})|^2$ by Equation (10). The probability of an elementary scattering event is proportional to the scattering amplitude $B(\mathbf{K}, \mathbf{k})$ squared

$$W(\mathbf{k} \to \mathbf{K}) \propto |B(\mathbf{K},\mathbf{k})|^2. \tag{13}$$

There are different contributions to the function $F$ determined by Equation (11). The term with $q = 0$ corresponds to pure SW scattering processes without moment transfer from SW to DW, whereas the terms with a non-zero value of $q$ describe the momentum transfer from SW to DW. However, the processes with $q=0$ can be prohibited due to the energy conservation. If we account for the main terms in Equation (10) at $\xi \ll 1$, then the scattering probability is determined by the simple expression

$$W(\mathbf{k} \to \mathbf{K}) \propto \left[\left(K_y^2 - k_y^2\right)\right]^2. \tag{14}$$

We assume that the incident SW has a fixed wave vector directed perpendicularly to the DW plane $yOz$ (Figure 2), i.e. $k_y = 0$. Then, the intensity of the scattered SW is $W(\mathbf{k} \to \mathbf{K}) \propto K_y^4$. Let us calculate the intensity as function of $\varphi_s$, the angle of the vector $\mathbf{K}$ with respect to the DW normal direction $\mathbf{x}$. The function $W(\varphi_s)$ increases sharply with the scattering angle $\varphi_s$ increasing. The scattering probability $W(\varphi_s)$ for a given incident SW vector and frequency $(\mathbf{k},\omega_\mathbf{k})$ should be calculated separately for the emission (a) and absorption (b) processes. The particular form of the angular dependence $W(\varphi_s)$ is determined by the SW and Winter's magnon dispersion relations. We assume that the dispersion relations can be written in the form $\omega_\mathbf{k} = \omega(k^2)$, $\Omega_\kappa = \Omega(\kappa_y^2)$, where $k = |\mathbf{k}|$. This assumption means that the magnetostatic interaction leading to the anisotropy of the spectra in the **k**-space is accounted in a simplified form or neglected. The scattered SW wave vector is $K^2 = \omega^{-1}\left[\omega_{\mathbf{k}} \mp \Omega_{\mathbf{\kappa}}\right]$ due to energy conservation for the emission (−) and absorption (+) processes. Here $\omega^{-1}$ is the inverse function to $\omega_{\mathbf{k}} = \omega(k^2)$. The scattering angle $\varphi_s$ is determined by the relation $\sin^2 \varphi_s = K_y^2/K^2$ (see Figure 2). Therefore, we can get the following equation for $\kappa_z(\varphi_s)$:

$$\sin^2 \varphi_s = \frac{\kappa_y^2}{\omega^{-1}\left[\omega_{\mathbf{k}} \mp \Omega_{\mathbf{\kappa}}\right]}. \tag{15}$$

The scattering probability is $W(\mathbf{k} \to \mathbf{K}) \propto \kappa_y^4$ due to momentum conservation ($\kappa_y^2 = K_y^2$). Solving Equation (15) we get explicit form of the dependence $W(\varphi_s) \propto \kappa_y^4(\varphi_s)$ for any particular SW and DW frequency dispersion relations $\omega_{\mathbf{k}} = \omega(k^2)$, $\Omega_{\mathbf{\kappa}} = \Omega(\kappa_y^2)$. The simplest dispersion relations, corresponding to the case of out-of-plane magnetizations directions $\mathbf{M}_0(\Theta_0, \Phi_0)$ of the domains, are $\omega_{\mathbf{k}} = \omega_0 + Dk^2$, $\Omega_{\mathbf{\kappa}} = \Omega_0 + D\kappa_y^2$, where $\omega_0$, $\Omega_0$ are the frequency gaps due to magnetic anisotropy and dipolar interaction, and $D = 2\gamma A/M_s$ is the SW stiffness. The absorption process is allowed only for $\Omega_0 = 0$. For the quadratic dispersion relations, the scattering intensities expressed via the input wave vectors of SW (**k**) and Winter's magnons ($\kappa_y$) are ($0 \leq \varphi_s < \pi/2$)

$$\left|B(\varphi_s, k)\right|^2 = B_0^2 \left(k^2 - \frac{\Omega_0}{D}\right)^2 \frac{\sin^4 \varphi_s}{\left(1 + \sin^2 \varphi_s\right)^2}, \tag{16a}$$

$$\left|B(\varphi_s, k, \kappa_y)\right|^2 = \left(\frac{\pi}{2}\right)^2 B_0^2 \left(k^2 + \kappa_y^2\right)^2 \sin^4 \varphi_s, \tag{16b}$$

for the emission and absorption processes, respectively. Here we introduced the parameter $B_0 = 2AV_0/\pi\sqrt{N_w}$.

For the case of in-plane magnetizations directions $\mathbf{M}_0(\Theta_0, \Phi_0)$ of the domains (Figure 1), the dispersion relations are more complicated: $\omega_{\mathbf{k}}^2 = (\omega_0 + Dk^2)(\omega_0' + Dk^2)$ and $\Omega_{\mathbf{\kappa}}^2 = (\Omega_0 + D\kappa_y^2)(\Omega_0' + D\kappa_y^2)$, where $\omega_0, \omega_0', \Omega_0, \Omega_0'$ are the contributions to the SW frequency due to magnetic anisotropy and dipolar interaction. However, solving Equation (15) we can get an explicit form of the dependence $W(\varphi_s) \propto \kappa_y^4(\varphi_s)$ for the in-plane magnetized magnetic element.

The probability of the SW scattering for emission and absorption processes $W(\varphi_s)$ essentially increases with the scattering angle $\varphi_s$ increase (see Figure 3). We note that the angular dependences $W(\varphi_s)$ are different for the Winter's magnon emission and absorption processes — see Equation (16). This difference originates from the energy and momentum conservations laws in course of these elementary scattering events. It is more difficult to satisfy the energy conservation for the absorption processes. Therefore, for these processes DW magnons should have low energy (no frequency gap). There is no momentum trasfer to the DW as whole for the absorption processes ($q = 0$) and the increase of the scattering angle $\varphi_s$ (the z-component of the SW wave vector) facilitates the momentum conservation. That is the reason for stronger dependence of the scattering amplitude on $\varphi_s$ for the absorption processes.

The angular dependence $W(\varphi_s)$ is stronger for the magnon absorption process that allows distinguishing these elementary magnon emission/absorption processes experimentally.

The scattering probability is determined by the squared amplitude of the SW scattering $W(\mathbf{k} \to \mathbf{K}) \propto |B(\mathbf{K},\mathbf{k})|^2$, where $|B(\mathbf{K},\mathbf{k})|^2 = (2/\pi)^2 A^2 V_0^2 K_y^4 / N_w \propto A^2 K_y^4$ is determined by the exchange interaction and increases with the increasing the input SW wave vector lengths (if the SW wave length $\lambda_{SW} \gg \Delta$). We note that the three-magnon interaction term similar to one in Equation (9) and (10) was very recently suggested by Zhang et al.[35] analyzing a formal decomposition of the Hamiltonian on the SW amplitudes. The principal difference of both approaches consists in the different mechanisms of three-magnon interactions. In this paper we investigate the exchange mechanism of nonlinear SW interaction and amplitude of this process is proportional to the exchange stiffness constant. Whereas in Ref.[35], only magnetic anisotropy was considered as origin of the three-magnon interaction.

The approach to the SW inelastic scattering used above is based on the interaction of the SWs with the dynamical emergent electromagnetic field generated by the moving inhomogeneous background magnetization texture (in our case, oscillating DW). Such emergent field results, in particular, in the skyrmion/vortex Hall effect and the topological Hall effect detected experimentally in magnetic films and stripes.[34] The non-linear SW-DW interaction can be also considered as origin of an addition damping of the initial propagating SW (**k**, ω**k**). There is no need to excite the Winter´s magnons for the absorption process because of a finite population of the magnon states at finite temperature. Such interpretation of the SW-DW interaction demands introducing the magnon population numbers calculating the matrix elements of the interaction Hamiltonian (9) and will be considered elsewhere.

To conclude, we calculated the inelastic spin wave scattering by vibrations of the Bloch domain wall (Winter's magnons) due to cubic nonlinearity in the magnetic subsystems. The obtained results are quite general. They are also valid for the Neel domain walls ($\Phi_0 = 0, \pi$) in thin films with an uniaxial magnetic anisotropy. The particular forms of the spin wave spectra of the magnetic film and domain wall (the equilibrium magnetization direction in the domains, magnetic anisotropy and dipolar interaction) result in specific angular dependences of the spin wave scattering probability. These angular dependences are different for the emission and absorption of the Winter's magnons by propagating spin waves. The scattered spin wave modulated by domain wall oscillations in ferromagnetic materials can be registered with the methods of microwave or micro-BLS spectroscopy.[36]

We believe that the considered phenomenon can serve as an additional tool to study the spin wave and domain wall dynamical properties in thin magnetic films and other systems, for example, in magnetic stripes and multilayers. Exploiting the flexure domain wall oscillations opens a new way to manipulate the high-frequency spin wave propagation on the nanoscale redirecting them by the large angles up to $90^0$.

**Acknowledgements**
This research is supported via the EU's Horizon 2020 Research and Innovation Program under Marie Skłodowska-Curie Grant Agreement No. 644348 (MagIC). Y.S.D. acknowledges support by the Ministry of Education and Science of Russia (Project No. 3.7614.2017/9.10).

N.N.D. acknowledges support by the Ministry of Education and Science of Russia (Project No. 14.Z50.31.0015 and the State Contract 16.2773.2017/4.6). K.Y.G. acknowledges support by IKERBASQUE (the Basque Foundation for Science) and the Spanish MINECO grant FIS2016-78591-C3-3-R.

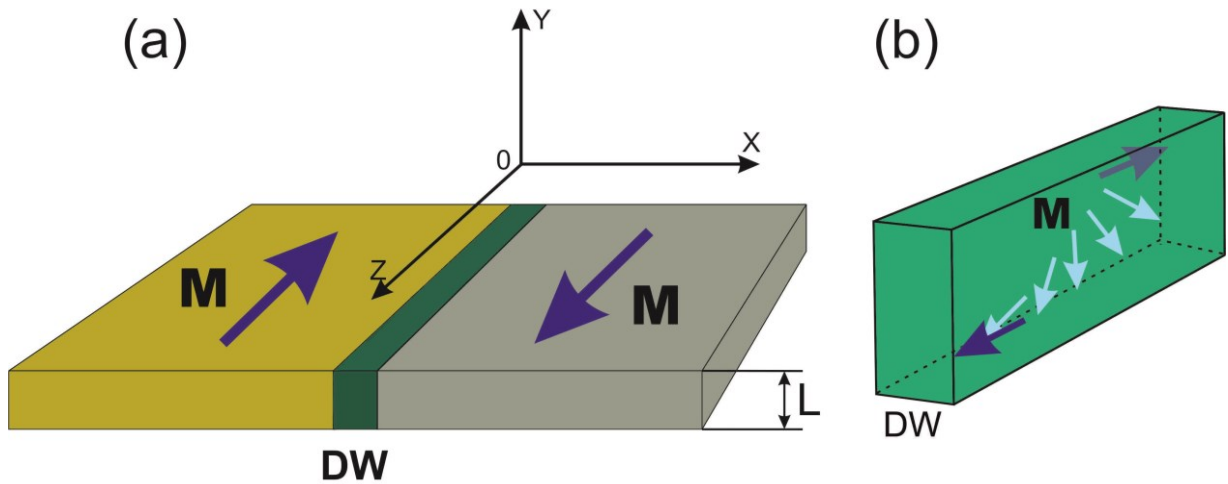

**Figure 1.** Sketch of the thin film element (a) with the Bloch DW (b) in the middle (marked by deep green color). The magnetization vectors are in-plane (as shown at Figure 1(a)) in the left and right domains, and out-plane in the DW(Figure 1(b)).

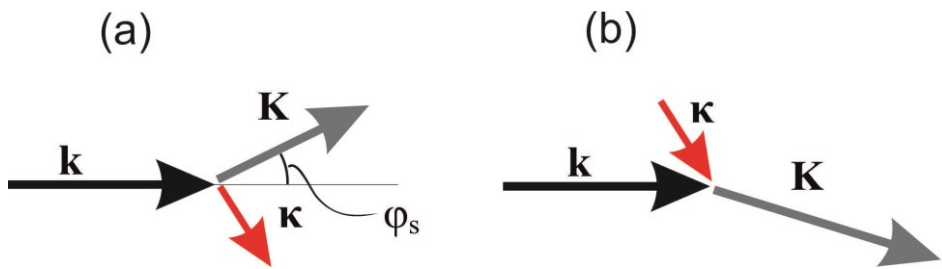

**Figure 2.** Inelastic scattering of the SW with the wave vector/frequency $(\mathbf{k}, \omega_\mathbf{k})$ to the SW with $(\mathbf{K}, \omega_\mathbf{K})$ due to emission (a) or absorption (b) of the DW localized excitation with $(\boldsymbol{\kappa}, \Omega_\boldsymbol{\kappa})$. The DW excitation wave vector is $\boldsymbol{\kappa} = (q, \kappa_y, 0)$, where $\kappa_y$ component corresponds to the excitation of the Winter's magnon with the frequency $\Omega(\kappa_y)$, and $q = k_x - K_x$ describes the momentum transfer from the incident SW to DW as a whole.

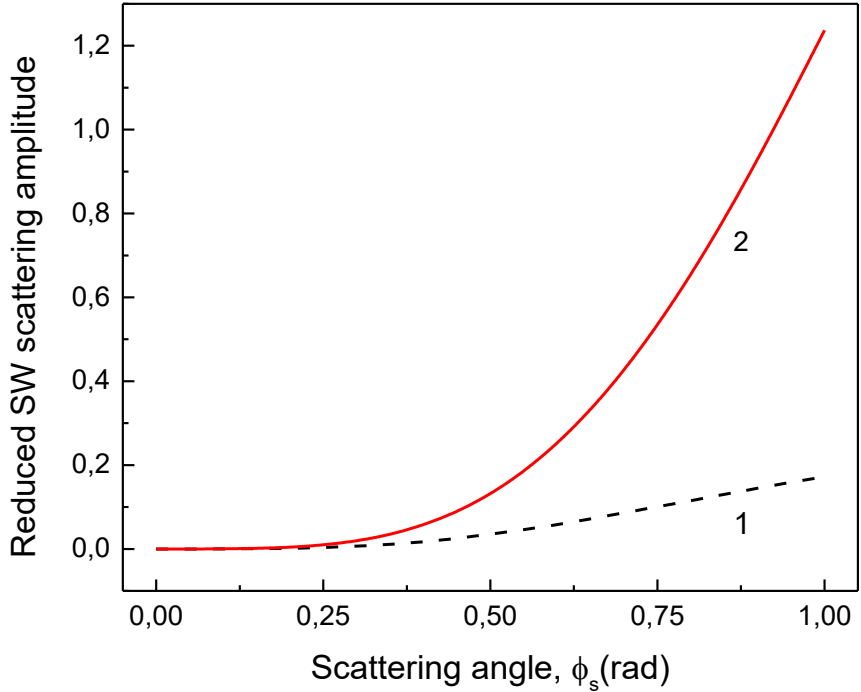

**Figure 3.** Reduced scattering amplitude $|B(\mathbf{K}, \mathbf{k})|^2 / B_0^2 k^4$ vs. the SW scattering angle $\varphi_s$ according to Equation (16): 1 – the Winter's magnon emission process **(16a)**, 2 – the Winter's magnon absorption process **(16b)**. The incident SW $(\mathbf{k}, \omega_\mathbf{k})$ propagates normally to the DW plane $yOz$, $\mathbf{k} = (k_x, 0)$, $\kappa_y = 0.2 k_x$.

**The table of contents**

**The calculations of the inelastic spin wave scattering** by flexure vibrations of the Bloch domain wall (Winter's magnons) in thin magnetic films are presented. The calculated angular dependence of the spin wave scattering probability is essentially stronger for the Winter's magnon absorption processes that allow distinguishing the elementary magnon emission/absorption processes experimentally.

**Keyword: spin waves**

N. N. Dadoenkova, Y. S. Dadoenkova, I. L. Lyubchanskii, M. Krawczyk, K. Y. Guslienko*

**Inelastic spin-wave scattering by Bloch domain wall flexure oscillations**

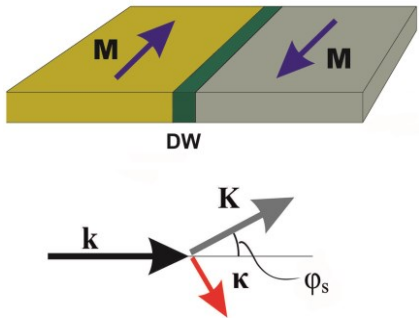